\documentstyle[floats,epsfig,aps,prb]{revtex}

\newcommand{\be}{\begin{equation}}
\newcommand{\ee}{\end{equation}}
\newcommand{\bea}{\begin{eqnarray}}
\newcommand{\eea}{\end{eqnarray}}
\newcommand{\om}{\omega}

\newcommand{\al}{\alpha}
\newcommand{\dl}{\delta}
\newcommand{\Dl}{\Delta}

\newcommand{\prll}{\parallel}

\newcommand{\th}{\theta}

\begin{document}

\title{\bf Adsorbate aggregation and relaxation of low-frequency vibrations}
\author{M. V. Pykhtin}
\address{Department of Physics and Astronomy and Center for Simulational Physics, 
University of Georgia, Athens, Georgia 30602-2451}
\author{Andrew M. Rappe}
\address{Department of Chemistry and Laboratory for Research on the Structure 
of Matter, University of Pennsylvania, Philadelphia, Pennsylvania 19104-6323}
\author{Steven P. Lewis}
\address{Department of Physics and Astronomy and Center for Simulational Physics, 
University of Georgia, Athens, Georgia 30602-2451}
\date{\today}
\maketitle
\draft

\begin{abstract}

\end{abstract}
We present a study of resonant vibrational coupling between adsorbates 
and an elastic substrate at low macroscopic coverages. In the first part of 
the paper we consider the situation  
when adsorbates form aggregates with high local coverage. 
Based upon our previously published theory, we derive formulas describing 
the damping rate of adsorbate vibrations for two cases of such aggregation: 
(i) adsorbates attached to step edges and 
(ii) adsorbates forming two-dimensional islands. 
We have shown that damping is governed by local coverage. Particularly, 
for a wide range of resonant frequencies, the damping rate of adsorbates 
forming well separated islands is described by the damping rate formula 
for a periodic overlayer with the coverage equal to the local coverage in the island. 
The second part of the paper is devoted to facilitating the evaluation of 
damping rates for a disordered overlayer. The formula describing the damping rate 
involves the parameter $\beta$ which is related to the local density of 
phonon states at the substrate surface and does not allow a closed-form 
representation. For substrates of isotropic and cubic symmetries, we have 
developed a good analytical approximation to this parameter. For a vast 
majority of cubic substrates the difference between the analytical approximation 
and numerical calculation does not exceed 4\%.



\section{Introduction}
\label{sec_intro}

Molecules or atoms adsorbed on surfaces of materials can oscillate 
about the adsorption bond. If this motion corresponds, in the absence 
of the adsorption bond, to free translation parallel to the surface, 
then this mode is called a frustrated translation (FT). 
It plays an important role in such dynamical processes as reactivity, 
surface diffusion, and desorption. For many systems, especially with 
a metallic substrate, the frequency of the FT mode is so low that 
it readily couples to low-energy excitations in the substrate, which in 
the case of metals include both phonons and electron-hole pair excitations. 
These bulk excitations provide decay channels for FT modes, which thus 
acquire a finite lifetime $\tau$ and become resonances. Associated with 
this lifetime is the resonance width, or damping rate, $\gamma = 1 / \tau$. 
In many cases the electronic contribution to the damping rate is small 
compared with the phononic contribution.\cite{Pers88} 

We recently developed a general theory of phonon-mediated FT relaxation 
that can be applied to any adsorbate overlayer pattern. Using this theory 
we derived simple FT damping rate laws for three important classes of 
adsorbate patterns: an isolated adsorbate, an ordered overlayer, and 
a randomly disordered overlayer. Although there are no experimental results 
to compare to for isolated adsorbates, the agreement with experiment for 
ordered and disordered overlayers is excellent.\cite{OurPRL} 

However, these three structural models do not cover all possible experimental 
conditions. Depending on the interactions of the adsorbates with the substrate 
and with each other, their distribution on the surface can differ strongly 
from the periodic or random uncorrelated cases.
Under certain conditions the adsorbates may gather into aggregates 
even at very low coverages. In this paper we consider two possible examples 
of such adsorbate aggregation: (i) adsorbates accumulating along the edges of 
surface steps, and (ii) adsorbates forming two-dimensional (2D) islands. We assume 
that the {\it average} coverage is low enough that these aggregates are 
well separated from each other and can be treated as isolated. For each of 
these cases we derive a corresponding damping rate formula from our general 
result. 

Another goal of this paper is to facilitate the evaluation of damping rates 
for a disordered overlayer. The main difficulty 
here is that one needs to integrate the Green's function for a semi-infinite 
elastic medium over the in-plane wavevector at a dense array of frequencies. 
Regrettably, no closed-form expression for this integral exists, and it must 
be computed numerically. 
In this paper we present a good analytical approximation of this integral 
for elastic media of isotropic and cubic symmetries. 
We have tested our approximation formula for 88 different 
materials with cubic symmetry, comparing it against well-converged 
numerical calculations. The difference between the analytical approximation and 
the reference numerical values never exceeds 6\%, and for a vast majority of 
the materials is under 4\%.

The remainder of this paper is organized as follows. In Section~\ref{sec_basic} 
we briefly review our theory of phonon-mediated FT damping. For details of 
the derivation, the reader is referred to Ref.~\protect\onlinecite{OurPRL}. 
This theory is then applied in Sections~\ref{sec_steps} and \ref{sec_islands} to the 
cases of adsorbates aggregated along step edges and in islands, respectively. 
In Section~\ref{sec_approx} we present our approximation formula for the 
$\vec{k}$-space integral of the substrate Green's function, thus facilitating 
the evaluation of the damping rate of a disordered overlayer. Finally, 
in Section~\ref{sec_concl} we summarize and draw conclusions.


\section{Review of theory}
\label{sec_basic}
The phonon contribution to the damping of low-frequency adsorbate 
vibrations was considered earlier by Persson and Ryberg~\cite{Pers85} (PR)  
and by Hall, Mills, and Black~\cite{Mills} (HMB).
Persson and Ryberg considered a single, isolated adsorbate coupled to a 
semi-infinite isotropic elastic substrate. Since direct adsorbate-adsorbate 
interactions are weak for all but the densest coverages, this case was 
presumed applicable for most experimental conditions.  Their model gives a 
damping rate law that varies with the vibrational frequency as $\omega_0^4$. 
However when applied to various experimental systems, the PR model predicts 
damping rates that are much smaller than the measured values.  Furthermore, 
the PR model has no explicit coverage dependence; coverage enters the 
damping-rate law only implicitly through its effect on the resonance frequency. 
The applicability of the model is limited by extremely low coverage ($\le 1\%$). 
Hall, Mills, and Black considered a periodic overlayer of rare-gas 
molecules physisorbed at an isotropic substrate. The damping rate implied by 
their result is proportional to the coverage and to the square of the resonant 
frequency. The periodic distribution of the adsorbates over the surface implied 
by the HMB model generally occurs at high coverage ($\ge 25\%$). 
In general, most adsorbate systems have structures that differ from both 
of these special cases.

In our recent work~\cite{OurPRL} we developed a general theory for 
a completely arbitrary overlayer structure. We considered an array of $N$ 
non-interacting adsorbates of mass $m$ located at arbitrary positions 
$\{\vec{R}_a\}$ at the surface of an anisotropic, semi-infinite elastic 
substrate of mass density $\rho$ and elastic modulus tensor $C_{ijkl}$, 
occupying the half-space $z<0$. The adsorbates are coupled to the substrate 
by harmonic springs of frequency $\omega_0$ and are allowed to oscillate 
along one of the Cartesian axes $\alpha$. The adsorbates do not interact 
with each other directly. We showed that the damping rate for this system 
is described by 
\be
\gamma = {m \over \rho}\, \om_0^2 \, \int \frac{d^2k}{(2 \pi)^2} 
\left[\om_0\,{\rm Im}\,D_{\al\al}(\vec{k},\om_0)\right] S(\vec{k}) \equiv 
{m \over \rho}\, \om_0^2 \,I(\om_0)\;,
\label{gam0}
\ee
where $D_{ij}(\vec{k},\om)$ is the in-plane Fourier transform 
of the substrate elastic Green's function evaluated at the surface ($z=z\,'=0$), 
and $S(\vec{k})$ is the structure factor defined as
\begin{equation}
S(\vec{k}) = \frac{1}{N}\left\langle \sum_{a,b}
\exp[-i\,\vec{k}\cdot (\vec{R}_a-\vec{R}_b)] \right\rangle\;,
\label{s0}
\end{equation}
where the angle brackets denote the appropriate ensemble average. 
The integration in Eq.~(\ref{gam0}) is taken over the bare-surface Brillouin zone. 
The integral $I(\om)$ in Eq.~(\ref{gam0}) is a convolution of two 
quantities, one of which (inside the square brackets) describes all the substrate 
properties, while $S(\vec{k})$ provides all the information on how 
the adsorbates are distributed over the surface. 
Except for a unitless factor of order unity, the quantity in the square brackets is 
equal to the $\vec{k}$-resolved density of elastic modes at the substrate surface 
for frequency $\om_0$. 
Any possible explicit dependence of the damping rate upon the coverage or 
the structural order comes from the structure factor $S(\vec{k})$. 

We previously applied this general result to three different structure models: 
an isolated adsorbate, a periodic array of adsorbates (appropriate at 
high coverages), and a random array of adsorbates (appropriate at low 
coverages). For each of these models, the structure factor has 
a simple algebraic form. Their substitution into Eq.~(\ref{gam0}) 
leads to 
\be
      \gamma =  {m \over \rho}\, \om_0^4 \,\beta
\label{gam1}
\ee
for an isolated adsorbate,
\be
      \gamma = {m \over \rho}\, \om_0^2 \, {\theta \over A_0 c_\perp^{(\al)}}
\label{gam2}
\ee
for a periodic overlayer,\cite{Hansen,OurJCP} and 
\be
      \gamma = (1-\theta)\,{m \over \rho}\, \om_0^4 \,\beta + 
         {m \over \rho}\, \om_0^2 \, {\theta \over A_0 c_\perp^{(\al)}}
\label{gam3}
\ee
for a random overlayer,\cite{correct}
where $\theta$ is the coverage, $A_0$ is the area per surface site, 
$c_\perp^{(\al)}$ is the speed of an $\al$-polarized sound wave propagating 
perpendicular to the surface, and $\beta$ is defined according to
\be
\beta = {1 \over \om} \, \int \frac{d^2k}{(2 \pi)^2} \;{\rm Im}\,D_{\al\al}(\vec{k},\om) 
\label{beta0}
\ee
The integral in Eq.~(\ref{beta0}) is proportional to the density of $\al$-polarized 
states divided by frequency. For a 3D elastic medium the density of $\al$-polarized 
states is proportional to $\om^2$, and thus $\beta$ is independent of frequency $\om$. 
The evaluation of this parameter is discussed in Section~\ref{sec_approx}. 
The damping rates given by Eqs.~(\ref{gam1}) and (\ref{gam2}) are just the 
anisotropic versions of the PR~\cite{Pers85} and HMB~\cite{Mills} results, respectively. 
No experimental results are available for an isolated adsorbate. 
However, we compared the other two models with 
experiments on CO on Cu(001) by applying Eq.~(\ref{gam2}) to the case of an ordered 
overlayer at 50\% coverage,\cite{GSHC} and Eq.~(\ref{gam3}) to the case of a 
disordered overlayer at 3\% coverage.\cite{GHT} 
Both models predict the lifetimes in excellent agreement with the measured values. 
Equation (\ref{gam3}) was independently derived later 
by Persson {\it et al.}~\cite{Pers99} for the case of an isotropic substrate. 
They also considered the case of an isolated adsorbate of finite size.


\section{Adsorbates aligned along step edges}
\label{sec_steps}

In this Section we study the case of adsorbates located at step edges. 
This geometry is important since steps are always present at surfaces 
and they generally attract adsorbates to their edges.\cite{Torri} While the 
macroscopic coverage is low for this case, the local coverage is high.

In our analysis we ignore the influence of the steps on the substrate 
Green's function. The only effect of steps that we consider is to set 
the distribution of adsorbates. We assume that the adsorbates form 
one-dimensional (1D), locally periodic arrays. Starting from 
Eq.~(\ref{gam0}) we derive the damping rate for an infinite, 
1D periodic chain of adsorbates, and then we extend the derivation for 
a finite, locally periodic chain.

\subsection{One-dimensional infinite chain}

We assume that the adsorbates are arranged in an infinite, 1D, periodic 
array along the $x$ axis. The adsorbate positions are given by 
$\vec{R}_a = (X_a,0)$, and the nearest-neighbor distance is $d$. 
The 1D coverage of the step edge can be defined as 
$\th_{1D} = a_0 / d$, where $a_0$ is the nearest-neighbor distance 
between the substrate atoms along the step edge.
Then the structure factor $S(\vec{k})$ takes the form
\be
S(\vec{k}) = {2 \pi \over d} \, \sum_{n} \, \dl(k_x - G_n) \;,
\label{s1}
\ee
where $\{G_n\}$ is the set of 1D reciprocal vectors defined as 
$G_n = 2 \pi n / d$. 
Substitution of Eq.~(\ref{s1}) into Eq.~(\ref{gam0}) leads to
\be
\gamma = {m \over \rho}\, \om_0^3 \,{1 \over d} \int \frac{dk_y}{2 \pi} \,
\sum_{n} \left[{\rm Im}\,D_{\al\al}(G_n,k_y;\om_0)\right] \;.
\label{gam5}
\ee

Since there are no resonant phonon states available in the substrate for in-plane 
wavevectors $|G_n| > \om_0 / c_R$, where $c_R$ is the speed of the Rayleigh 
wave, the imaginary part of the Green's function in Eq.~(\ref{gam5}) vanishes
for these values of $G_n$, and thus these terms do not contribute to the sum.
Similar to the case of a two-dimensional (2D) periodic overlayer discussed 
in detail in our earlier work,\cite{OurJCP} it turns out that for all 
1D coverages presently accessible to experiment ($\th_{1D} > \th_{1D}^{(c)} 
\approx 10\%$), only the term with $G_n = 0$ contributes to the damping rate. 
Moreover, the integral of this term in Eq.~(\ref{gam5}) does not depend on 
frequency $\om_0$, and we can define a constant $\zeta_\al$ according to
\be
\zeta_\al = \int \frac{dk_y}{2 \pi} \,
\left[{\rm Im}\,D_{\al\al}(0,k_y;\om_0)\right] \;.
\label{eta0}
\ee
It turns out that $\zeta_x$ (vibration parallel to the step) differs 
significantly from $\zeta_y$ (vibration perpendicular to the step). 
However, for substrates with isotropic or cubic symmetry, these 
values, while different from each other, are independent of the 
orientation of the step. It is, therefore, appropriate 
to use the notation $\zeta_\prll$ for oscillations along the step and 
$\zeta_\perp$ for oscillations perpendicular to the step but still 
in the plane. We can now write the damping rate formula compactly as
\be
\gamma_{\,\prll,\perp} = {m \over \rho}\,\om_0^3 \, 
{\zeta_{\,\prll,\perp} \over d}\;.
\label{gam6}
\ee
Thus, the oscillations along the step are damped at a different rate 
than the oscillations perpendicular to the step. Typically, the damping 
rate is higher for vibrations along the step. 

Experimentally measured damping rates can be related to Eq.~(\ref{gam6}) in 
two possible ways. If the frequencies for both vibrational 
modes are close, then the measured damping rate will correspond to 
an average of $\gamma_\prll$ and $\gamma_\perp$. Otherwise, both modes 
can be resolved experimentally, each exhibiting its own damping rate.

\subsection{One-dimensional finite chain}

\begin{figure}
\epsfysize=4in
\centerline{\epsfbox[18 144 592 718]{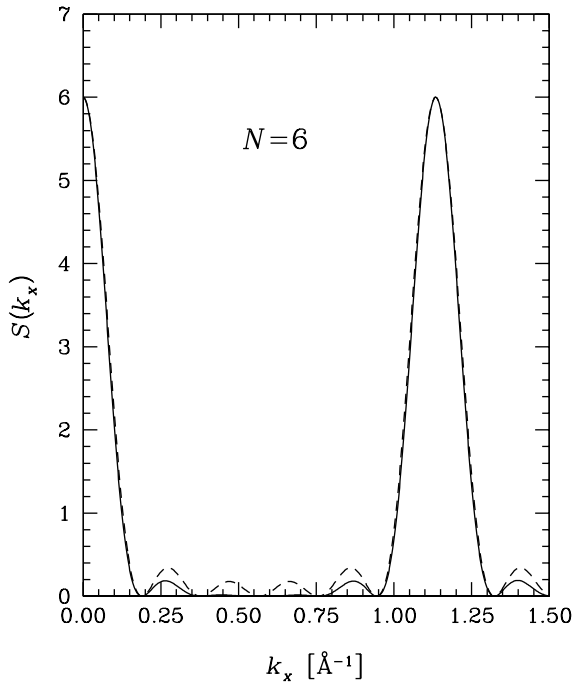}}
\caption{\label{sdiff} The structure factor $S(k_x)$ for a chain of 6 atoms 
separated by $d = 5.54$~\AA. The solid (dashed) curve was computed according to 
Eq.~(\ref{s2}) (Eq.~(\ref{s3})).
}
\end{figure}

Here we consider a 
finite, locally periodic chain of $N$ adsorbates along one of the step 
edges. As before, the $x$-axis is along the step, and the nearest-neighbor 
distance in the chain is $d$. The total length of the chain is $L = N\,d$.  
The structure factor corresponding to this configuration is given by
\be
S(\vec{k}) = N \left[ \sum_n \frac{\sin \left[ (k_x - G_n)\,L/2 \right]}
{\left[(k_x - G_n)\,L/2\right]}\right]^2 \;.
\label{s2}
\ee
Most of the weight of this function is in the peaks centered at 
the reciprocal vectors $G_n = 2 \pi n / d$. Even at relatively 
small values of $N$ the cross terms in Eq.~(\ref{s2}) may be neglected, 
and the square of the sum can be approximated by a sum of squares
\be
S(\vec{k}) = N \sum_n \left[ \frac{\sin \left[ (k_x - G_n)\,L/2 \right]}
{\left[(k_x - G_n)\,L/2\right]} \right]^2 \;.
\label{s3}
\ee
Equations (\ref{s2}) and (\ref{s3}) are compared for a chain of $N=6$ adsorbates 
separated by $d = 5.54$~\AA (twice the nearest-neighbor distance for platinum) 
in Fig.~\ref{sdiff}. The solid line corresponds to the structure factor computed 
according to Eq.~(\ref{s2}), while the dashed line represents $S(\vec{k})$
given by Eq.~(\ref{s3}). As $N$ increases, the main peaks become narrower 
(the width of the peaks is approximately $2\pi/L$), and the difference 
between the two formulas decreases. In the limit $N \to \infty$, both 
expressions take the form of Eq.~(\ref{s1}). 

Finally, we show that the damping rate for a finite chain 
of the adsorbates can be computed approximately from Eq.~(\ref{gam6}) for 
large enough $N$ (where ``large enough'' may be as small as 6). 
First, we assume that all the weight in the structure factor comes from 
the main peaks. Then, repeating the argument we gave for the infinite chain 
we conclude that if $\;\om_0 / c_R < 2\pi\,(1 - 1/N)\, /\, d\,$, only the peak 
centered at $G=0$ contributes to the integral in Eq.~(\ref{gam0}), and 
the damping rate becomes
\be
\gamma = {m \over \rho}\, \om_0^3 \,\int \frac{dk_x}{2 \pi} \,
N\,\left[\frac{\sin \left(k_x L/2 \right)}{(k_x L/2)} \right]^2
\int \frac{dk_y}{2 \pi}\left[{\rm Im}\,
D_{\al\al}(\vec{k},\om_0)\right] \;.
\label{gam7}
\ee
If the peak is narrow enough, the imaginary part of the Green's function 
does not change significantly over its width (the region $k_x < 2\pi/L$). 
Thus, for each value of $k_y$ the Green's function can be replaced by its 
value at $k_x = 0$. The integral over $k_x$ can now be carried out 
explicitly, and Eq.~(\ref{gam6}) follows.
 

\section{Adsorbates forming islands} 
\label{sec_islands}

In this Section we analyze the damping of the FT mode when adsorbates 
form 2D islands at the substrate surface. As with 1D step-edge aggregation, 
this case corresponds to low macroscopic coverage, but high local coverage. 
We assume that each island is a locally periodic array of $N$ adsorbates 
with local coverage $\th_{\rm loc}$ and area $A$. We also assume that 
the distance between islands is large enough that the islands can be treated 
as isolated. 

Unlike in the previous Section, there is no need here for considering 
the case of an infinite island, because 
an infinite island is the same as a periodic overlayer with 
$\th_{\rm loc} = \th$, and its FT damping rate is given by Eq.~(\ref{gam2}). 
The problem of a finite island appears to be more complicated than that 
of a finite 1D chain at a step edge. The reason is that, apart from its 
local coverage and size, the island is also characterized by its shape. 
It is convenient to represent the shape of the island by a function 
$f(\vec{r})$ defined as
\be
f(\vec{r}) = \left\{
      \begin{array}{ll}
	1 \:\:\:\: \mbox{if $\vec{r}$ is inside island} \\
	0 \:\:\:\: \mbox{otherwise}
      \end{array}
\right.
\label{f}
\ee
It is straightforward to show that the structure factor can be expressed 
in terms of the Fourier transform $f(\vec{k})$ of this function as 
\be
S(\vec{k}) = {N \over A^2}\, \sum_{\vec{G},\vec{G}\,'} f(\vec{k}-\vec{G})\, 
f(\vec{k}-\vec{G}\,') \;,
\label{s4}
\ee
where summation is over reciprocal vectors defined by the local periodicity 
of the adsorbate distribution inside the island.
 
We further assume that the island is a convex set with similar dimensions 
in all directions (i.e. its width is of the same order of magnitude 
as its length). If the number of adsorbates in the island is not too small 
(the meaning of which is discussed below), then 
the function $f(\vec{k})$ will have a strong peak at $\vec{k}=0$, with most 
of the function's weight within the peak. Analytical examples of this function 
include  
\be
f_{\rm rect}(\vec{k}) = A \,\frac{\sin \left[ k_x a/2 \right]}{\left[k_x a/2\right]}\,
\frac{\sin \left[ k_y b/2 \right]}{\left[k_y b/2\right]} \;.
\label{frect}
\ee
for a rectangular island with 
dimensions $a$ along the $x$-axis and $b$ along the $y$-axis, and 
\be
f_{\rm circ}(\vec{k}) = 2 A \, \frac{J_1(|\vec{k}| R)}{|\vec{k}| R}
\label{fcirc}
\ee
for a circular island of radius $R$, where $J_1(x)$ is the first-order 
Bessel function.

We now follow the arguments of the 1D case. First, we neglect 
the cross-terms in Eq.~(\ref{s4}) and write it as 
\be
S(\vec{k}) = {N \over A^2} \sum_{\vec{G}} \left[f(\vec{k}-\vec{G})\right]^2  \;.
\label{s5}
\ee
Then, we assume that all the weight in the structure factor comes from 
the main peaks. If the resonant frequency of the FT mode $\om_0$ is small 
enough that $\;\om_0 / c_R < G_1 - \Dl\,$, where $G_1$ is the 
magnitude of the smallest reciprocal vector, and $\Dl$ is the maximal width of 
the main peak in $S(\vec{k})$, then only the $\vec{G}=0$ term 
will contribute to the integral in Eq.~(\ref{gam0}). Thus this equation now 
takes the form
\be
\gamma = {m \over \rho}\, \om_0^2 \, \int \frac{d^2k}{(2 \pi)^2} 
\left[\om_0\,{\rm Im}\,D_{\al\al}(\vec{k},\om_0)\right] 
{N \over A^2}\,\left[f(\vec{k})\right]^2 \;.
\label{gam8}
\ee
If the the peak of $[f(\vec{k})]^2$ is narrow enough, the imaginary 
part of the Green's function 
does not change significantly over its width (i.e., the region $|\vec{k}|<\Dl$),  
and it can be replaced by its value at $\vec{k}=0$ given by 
$\;{\rm Im}\,D_{\al\al}(0,\om_0) = 1 / (c_\perp^{(\al)} \om_0)\,$, and
taken out of the integral. The remaining integral can be evaluated with 
the help of the Parseval's theorem as
\[
 \int \frac{d^2k}{(2 \pi)^2}\,{N \over A^2}\,\left[f(\vec{k})\right]^2 = 
 \int d^2r\,{N \over A^2}\,\left[f(\vec{r})\right]^2 = 
 {N \over A} = {\th_{\rm loc} \over A_0} \;,
\]
where, as before, $A_0$ is the area of the surface unit cell. Substituting 
this result into Eq.~(\ref{gam8}) we obtain
\be
      \gamma = {m \over \rho}\, \om_0^2 \, {\theta_{\rm loc} \over A_0 c_\perp^{(\al)}} \;,
\label{gam9}
\ee
which is exactly the damping rate of a periodic overlayer given by Eq.~(\ref{gam2}) 
with the total coverage $\th$ being replaced by the local coverage $\th_{\rm loc}$.

\begin{figure}[t]
\epsfysize=4in
\centerline{\epsfbox[18 144 592 718]{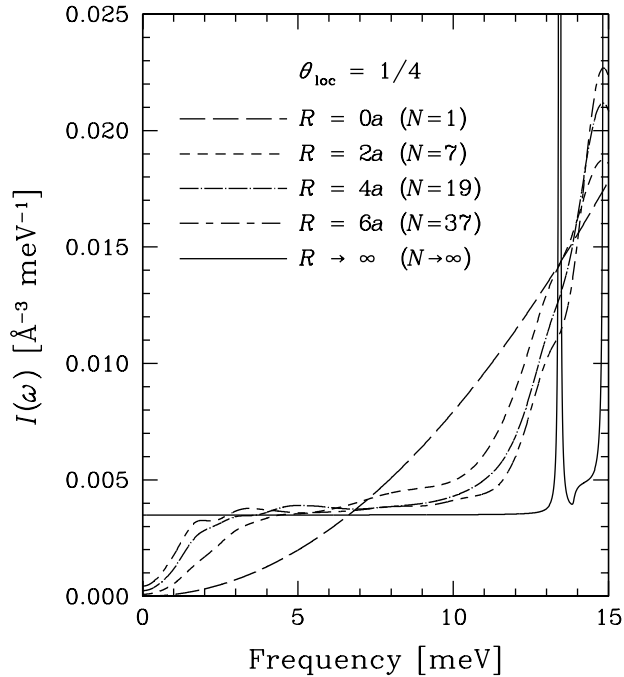}}
\caption{\label{islands} Function $I(\om)$ for an isolated adsorbate (dashed line), 
for an infinite periodic overlayer (solid line) at coverage $\th = 1/4$, and 
for one-shell (dotted line), two-shell (dash-dotted line), and three-shell 
(dash-double-dotted line) hexagonal islands at local coverage 
$\th_{\rm loc} = 1/4$. The substrate is Pt(111). In the inset, $R$ is the 
circumscribing radius of an island, $a$ is the nearest-neighbor distance 
between the substrate atoms, and $N$ is the number of adsorbates inside an island. 
The features at 13.3~meV are caused by coupling to Rayleigh waves at $G_1$.
}
\end{figure}

Finally, we need to determine for which values of $N$ the approximation given by 
Eq.~(\ref{gam9}) really works. We consider, as an example, an isolated island 
of hexagonal shape on the Pt(111) surface with local coverage $\th_{\rm loc} = 1/4$. 
This is about the lowest local coverage observed for islands of small molecular 
adsorbates. If the approximation given 
by Eq.~(\ref{gam9}) holds for this coverage, it will also hold for higher 
coverages. We have computed the integral $I(\om)$ in Eq.~(\ref{gam0}) for 
hexagonal islands of different sizes, and, in Fig.~\ref{islands}, plot 
$I(\om)$ vs. $\om$ for each island size.  The value of the integral for the 
two limiting cases of an isolated adsorbate and an infinite periodic overlayer are 
shown by the dashed and solid lines, respectively. The dotted, dash-dotted, and  
dash-double-dotted lines correspond to the cases of one, two, and three 
hexagonal shells of adsorbates around the center adsorbate, respectively. 
The inset to Fig.~\ref{islands} characterizes the islands by their 
circumscribing radius $R$, and the number of adsorbates, $N$, within the island. 

The behavior of $I(\om)$ shown in Fig.~\ref{islands} can readily be explained 
in terms of resonant phonon wavelengths at frequency $\om$. 
For small frequencies the phonon wavelengths are much greater than the 
island size, and the adsorbates in the island behave like a single isolated 
adsorbate of mass $N m$. Thus, for small frequencies we have 
$I(\om) = N \beta \om^2$ in accord with Eq.~(\ref{gam1}).\cite{note1} 
Mathematically, this result can be obtained from 
Eq.~(\ref{gam8}) by setting $f(\vec{k}) = f(0) = A$. 
Frequencies for which the resonant phonon wavelengths 
become comparable with the island size correspond to the cross-over region 
where this quadratic behavior changes. When the phonon wavelengths are smaller 
than the island size, but still significantly greater than the nearest-neighbor 
distance between the adsorbates, only phonons with in-plane wavevector component, 
$\vec{k}$, near zero couple strongly to the FT mode due to the $\vec{G} = 0$ 
peak in the structure factor given by Eq.~(\ref{s5}). Terms with 
$\vec{G} \neq 0$ have no corresponding resonant phonons at these frequencies. 
In this frequency region we expect $I(\om)$ to 
follow closely its limiting value of a full periodic overlayer. 
Finally, when $\om$ is large enough that the phonon wavelengths become 
comparable to the inter-adsorbate nearest-neighbor distance, 
non-zero $\vec{G}$-vectors in the structure factor become available channels 
for resonant phonon coupling. However, for realistic local 
coverages ($\th_{\rm loc} \ge 1/4$) these wavevectors lie at the boundary of 
the bare-surface Brillouin zone, where continuum models are no longer applicable.
For the case $\th_{\rm loc} = 1$, this second cross-over region does not exist, 
because all non-zero $\vec{G}$-vectors lie outside the surface Brillouin zone. 

From Fig.~\ref{islands} one can determine the range of resonant 
FT frequencies for which the approximation given by Eq.~(\ref{gam9}) holds. 
Even for the island with only one hexagonal shell ($N = 7$), there is 
such a frequency range: $I(\om)$ stays close to its limiting value of 
a periodic overlayer for a range of frequencies between about 3 and 7~meV. 
For larger islands this range expands to frequencies between 2 and 11~meV. 

We can analytically estimate this frequency range for an island of size $D$. 
The low-frequency threshold corresponds to a transverse wave with wavevector 
magnitude equal to the width, $\Dl$, of a peak in the structure factor, which 
can be estimated as $2 \pi / D$. The frequency of such a wave is 
$\om_{\rm low} = 2 \pi c_{\rm T} / D$. The high frequency threshold corresponds 
to a transverse wave with wavevector magnitude equal to $G_1 - \Dl$, which is 
approximately $2 \,(D/d - 1)\, \pi c_{\rm T} / D$, where $d$ is the inter-adsorbate 
nearest-neighbor distance, as before. The frequency of this wave is 
$\om_{\rm high} = (D/d - 1)\, \om_{\rm low}$. For a three-shell hexagonal island 
($N=37$) with $\th_{\rm loc} = 1/4$ on the Pt(111) surface, we have $d = 5.54$~\AA, 
$D/d = 6$, and $c_{\rm T} = 11$~\AA~meV, which leads to the estimates 
$\om_{\rm low} = 2.1$~meV and $\om_{\rm high} = 10.5$~meV. This frequency 
range is in excellent agreement with the corresponding plot in Fig.~\ref{islands}.
Thus, the example of a hexagonal island on the Pt(111) surface 
demonstrates that, when most of the adsorbates form widely separated islands, 
the damping rate can be computed from Eq.~(\ref{gam9}) for FT resonant 
frequencies within a broad range. 


\section{Approximation formula for $\beta$ parameter}
\label{sec_approx}

The main difficulty researchers may encounter in using 
Eq.~(\ref{gam3}) to determine the damping rate for low-coverage systems
is the necessity to compute the substrate Green's function and the 
integral in Eq.~(\ref{beta0}) numerically. In this Section we present 
an analytical approximation to the parameter $\beta$ defined by 
Eq.~(\ref{beta0}). The approximation is developed for the cases of 
systems with isotropic or cubic symmetry.

We start from the observation that $\beta$ must be a function of 
the substrate parameters of the model, namely, the mass density $\rho$ and the 
elastic constants $c_{ij}$. Generally, $i,j=1,...,6$, but we include 
only independent constants characteristic of a given symmetry. 
This function must obey certain global constraints, however. 
It follows from the equation of motion and 
the boundary conditions that any quantity depending on the model inputs 
may depend only on the ratios $c_{ij}/\rho$, which we denote as $k_{ij}$. 
Because $\beta$ does not depend on the frequency $\om$, 
one can show from the equations of motion that  $\beta$, as a function 
of density-normalized elastic constants $k_{ij}$, must satisfy a scaling 
property given by
\be
\beta(v k_{ij}) = v^{-3/2}\, \beta(k_{ij}) \;,
\label{scale}
\ee
where $v$ is an arbitrary number. This scaling property reduces the 
number of independent variables in the function $\beta(k_{ij})$ by one. 
We have found an excellent analytical approximation to this function 
for the cases of isotropic and cubic substrates.

\subsection{Isotropic substrate}

An isotropic medium is characterized by only two density-normalized 
elastic constants: $k_{11}$ and $k_{44}$. The transverse and the 
longitudinal speeds of sound are related to these constants by 
$c_{\rm T} = \sqrt{k_{44}}$ and $c_{\rm L} = \sqrt{k_{11}}$, respectively. 
Taking into account the scaling property given by Eq.~(\ref{scale}), we 
can express $\beta^{-2}$ as
\be
\beta^{-2} = k_{44}^3 \, g\left({k_{11} \over k_{44}}\right) \;,
\label{fitiso1}
\ee
where $g(x)$ is an unknown dimensionless function of one variable. 

\begin{figure}
\epsfysize=4in
\centerline{\epsfbox[18 144 592 718]{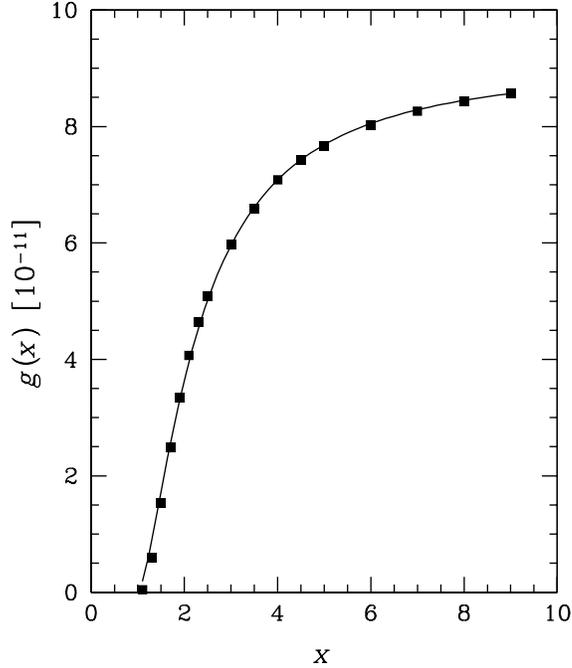}}
\caption{\label{fitiso} Fitting function $g(x)$ defined in Eq.~(\ref{fitiso1}), 
computed numerically (squares) and using the analytical approximation given 
in Eq.~(\ref{fitiso2}) (solid line).
}
\end{figure}

To elucidate the behavior of $g(x)$, we have computed $\beta$ numerically 
for a wide range of $k_{11}/k_{44}$. The result 
of these calculations is shown by squares in Fig.~\ref{fitiso}. 
From Fig.~\ref{fitiso} we obtain that (i) $g(x)$ goes to zero at $x=1$ 
as $(x-1)^\al$, where $1<\al<2$, and (ii) $g(x)$ asymptotically approaches 
a constant as $x \to \infty$. We find that the function 
\be
g(x) = P \, {(x-1)^{3/2} \over 3/2 + (x-1)^{3/2}}
\label{fitiso2}
\ee
almost perfectly approximates  the numerical data (solid line in Fig.~\ref{fitiso}). 
The parameter $P$ is dimensionless and has an optimal value of 
$P = 9.14 \times 10^{-11}$.   

\subsection{Cubic substrate}

The isotropic approximation is not correct for most single-crystal 
materials. We therefore extend our approximation formula for 
$\beta$ to describe substrate materials with cubic symmetry, since 
these are the most commonly used in experiments. Crystals with cubic 
symmetry have three independent density-normalized elastic constants: 
$k_{11}$, $k_{12}$, and $k_{44}$. The isotropic limit can be obtained from 
cubic symmetry by setting $k_{12} = k_{11} - 2 k_{44}$. Taking advantage 
of the scaling property given by Eq.~(\ref{scale}), we can write
\be
\beta^{-2} = k_{12}^3 \, h\left({k_{11} \over k_{12}}, 
{k_{44} \over k_{12}}\right) \;,
\label{fitcub1}
\ee
where $h(x,y)$ is an unknown function of two variables. Generally, the form 
of the function $h(x,y)$ may depend on the particular choice of the surface indices. 
However, numerical calculations of $\beta$ for different low-index surfaces 
of the same material show that $\beta$ is essentially independent 
of the surface orientation with respect to the crystallographic axes. 
Thus, the same function $h(x,y)$ works well for any low-index surface. 

First, we study how $h(x,y)$ depends on $x$ by computing $\beta$ numerically 
as a function of $k_{11}/k_{12}$ while keeping $k_{44}/k_{12}$ fixed. We find 
that $h(x,y)$ depends linearly on $x$ when $x \gg 1$ and goes to zero at $x=1$ 
as $(x-1)^{3/2}$. We choose the fitting function to have the form 
\be
h(x,y) = b(y) \, \frac{(x-1)^{3/2}}{y+(x-1)^{1/2}}
\label{fitcub2}
\ee
which satisfies these asymptotic requirements.

The next step is to find a fitting function for $b(y)$. This can be accomplished 
by requiring that $\beta$ for the cubic substrate given by 
Eq.~(\ref{fitcub1}) go to the isotropic limit described by 
Eq.~(\ref{fitiso1}) when $k_{12} = k_{11} - 2 k_{44}$. This requirement 
uniquely specifies that the function $b(y)$ has the form
\be
b(y) = {P \over 2}\, y^2 \, \left[ 1 + \left({y \over 2}\right)^{1/2} \right] \,
\frac{(y^{-1}+1)^{3/2}}{3/2+(y^{-1}+1)^{3/2}} \;,
\label{fitcub3}
\ee
where the parameter $P$ is the one determined for the isotropic case, 
namely $P = 9.14 \times 10^{-11}$. 

To check the quality of the approximation given by Eqs.~(\ref{fitcub1}) 
and (\ref{fitcub3}), we have computed $\beta$ according to this approximation 
and compared it with $\beta$ obtained by numerical calculations 
for 88 different materials with cubic symmetry. The error in 
the approximated value of $\beta$ never exceeds 6\% for this set of materials, 
and for the vast majority of cases it is below 4\%. Table~\ref{tab_beta} shows this 
comparison for several important substrate materials. It should be noted that 
a given error in $\beta$ results in a smaller error in the predicted damping rate, 
$\gamma$, due to the $\beta$-independent term in Eq.~(\ref{gam3}).
 
\begin{table}
\caption{\label{tab_beta}
Density-normalized elastic moduli ($10^{10}$~cm$^2$/s$^2$),
\protect\tablenote[1]{Mass density and elastic constants were 
taken from Ref.~\protect\onlinecite{elmod}.} $\beta_{\rm n}$ 
(computed numerically), $\beta_{\rm a}$ (computed using the analytical 
approximation), and $(\beta_{\rm a}-\beta_{\rm n})/\beta_{\rm n}$ 
for several cubic materials. All values of $\beta$ are given in 
$10^{-5}$~meV$^{-3}$\AA$^{-3}$.
}
\begin{tabular}{ldddddd}
Crystal & $k_{11}$ & $k_{12}$ & $k_{44}$ & $\beta_{\rm n}$ & 
$\beta_{\rm a}$ & $(\beta_{\rm a}-\beta_{\rm n})/\beta_{\rm n}$ \\
\hline
Al & 39.58 & 22.40 & 10.51 & 1.3635 & 1.3657 &  0.16\% \\
Cs & 1.25 & 1.04 & 0.75 & 242.75 & 231.03 & -4.83\% \\
Cu & 18.84 & 13.67 & 8.48 & 3.6321 & 3.5586 & -2.02\% \\
GaAs & 22.34 & 10.10 & 11.18 & 1.7750 & 1.7835 &  0.48\% \\
Ge & 24.16 & 9.08 & 12.55 & 1.4659 & 1.4608 & -0.35\% \\
Au & 9.98 & 8.45 & 2.18 & 23.549 & 22.361 & -5.05\% \\
Ir & 25.75 & 10.75 & 11.37 & 1.5337 & 1.5377 &  0.26\% \\
Fe & 28.73 & 17.80 & 14.74 & 1.4481 & 1.4374 & -0.74\% \\
MgO & 83.01 & 26.64 & 43.62 & 0.21876 & 0.22036 &  0.73\% \\
Ni & 27.85 & 17.38 & 13.94 & 1.5574 & 1.5439 & -0.87\% \\
Pd & 18.87 & 14.62 & 5.96 & 5.3473 & 5.1660 & -3.39\% \\
Pt & 16.13 & 11.66 & 3.56 & 7.9305 & 7.7329 & -2.49\% \\
Rb & 1.87 & 1.58 & 1.08 & 142.05 & 133.94 & -5.71\% \\
Si & 71.12 & 27.43 & 34.16 & 0.29820 & 0.30818 &  3.35\% \\
Ag & 11.81 & 8.92 & 4.39 & 9.0756 & 8.7958 & -3.08\% 
\end{tabular}
\end{table}


\section{Conclusion}
\label{sec_concl}

In this paper we have applied the general result of our continuum elastic 
model~\cite{OurPRL} of phonon-mediated adsorbate vibrational relaxation to 
derive analytic formulas for the FT damping rate when the adsorbates form 
aggregates on the surface. We have considered two possible forms of adsorbate 
aggregation: (i) when the adsorbates are aligned along step edges, and 
(ii) when the adsorbates form well separated islands. We have derived the 
expression for the damping rate of an infinite periodic chain of the 
adsorbates, and then shown that the damping rate of a finite locally periodic 
chain is described by the same expression for all chains above a small threshold 
length. For the case of an isolated island we have shown that, under mild 
restrictions on the island geometry, there is a wide range of possible FT 
resonant frequencies for which the damping rate is essentially the same as for 
a periodic overlayer with coverage equal to the island's local coverage. 

Finally, to facilitate the use of our previously derived expression for the 
FT damping rate at low coverage, we have developed an analytical formula 
which approximates the parameter $\beta$ appearing in the damping law. 
The formula is valid for both isotropic and cubic materials and achieves 
remarkable accuracy. We have tested the formula on 88 different cubic 
materials ({\it all} of whose elastic constants were readily available) 
and have found that our analytic approximation predicts $\beta$ with 
accuracy better than 6\%.


\section*{Acknowledgments}

Financial support for this research was provided by the National
Science Foundation under Grant No.~DMR 97-02514, the Air Force
Office of Scientific Research, Air Force Materiel Command, USAF, under
Grant No.~F49620-00-1-0170, and the University of Georgia Research
Foundation.  SPL acknowledges support of the Donors of
The Petroleum Research Fund, administered by the American Chemical
Society.  AMR acknowledges support of the Alfred P. Sloan Foundation.



\begin{thebibliography}{9}
\bibitem{note1} The lines in Fig.~\ref{islands} do not go 
exactly to zero at $\om = 0$ due to the approximation of the integral $I(\om)$
by a finite sum.
\end{thebibliography}
\end{document}